\documentclass[aps,twocolumn,superscriptaddress,showpacs,pre,floatfix]{revtex4-1}
\usepackage{times,amsmath,graphics,graphicx,makeidx,fancyhdr,geometry,color}
\usepackage{amssymb}
\usepackage{dcolumn}
\usepackage{bm}
\usepackage{hyperref}
\usepackage{amsfonts}
\def\be{\begin{equation}}
\def\ee{\end{equation}}
\def\ba{\begin{eqnarray}}
\def\ea{\end{eqnarray}}
\def\bc{\begin{center}}
\def\ec{\end{center}}
\def\nn{\nonumber}
\def\lsim{\raise0.3ex\hbox{$\;<$\kern-0.75em\raise-1.1ex\hbox{$\sim\;$}}}
\def\gsim{\raise0.3ex\hbox{$\;>$\kern-0.75em\raise-1.1ex\hbox{$\sim\;$}}}

\definecolor{azulin}{cmyk}{0.7,0.3,0.0,0.2}

\begin{document}

\title{A Fourth Neutrino and its Consequences on CP Asymmetries}
\author{D.~Delepine}\email{delepine@fisica.ugto.mx}
\author{C.~Lujan-Peschard}\email{carolup@fisica.ugto.mx}
\author{M.~Napsuciale}\email{mauro@fisica.ugto.mx}
\affiliation{Departamento de F\'{i}sica, Divisi\'on de Ciencias e Ingenier\'{i}a, Universidad de Guanajuato-Campus Le\'on, 
Lomas del Bosque 103, Fraccionamiento Lomas del Campestre, 37150, Le\'on, Guanajuato, M\'exico.}

\begin{abstract}
A general analysis of the consequences of including a fourth neutrino in the standard model matter content, 
on CP violating observables at neutrino oscillation experiments, is presented. Neutrino oscillations in vacuum and 
with matter effects are studied. For the former we update and generalize previous studies on CP asymmetries with 
an additional active neutrino using an updated fit of the PMNS mixing matrix. We study the 
values of the new CP violating phases which maximize the different CP asymmetries in T2K and MINOS-like setups 
aiming to elucidate if the new phases yield measurable effects in the most favorable case. We show that due to a 
combined effect of kinematics and unitarity it is possible to obtain an observable asymmetry 
in the survival channels without violating CPT.  For the MINOS-like setup, we find maximum 
asymmetries in vacuum of the order of $2\%$ and $4\%$ for the $\nu_\mu \to \nu_e$ and $\nu_e\to \nu_\tau$ channels respectively. 
For the T2K-like setup we obtain maximum asymmetries of the order of $6\%$ in the survival $\nu_\mu\to\nu_\mu$ channel.
Tree level matter effects enhance the former reaching asymmetries of the order of $10\%$ for  the $\nu_\mu \to \nu_e$ and 
$\nu_e\to\nu_\tau$ channels, while the $\nu_\mu\to \nu_\mu$ survival channel changes slightly depending on the mass hierarchy. Box diagrams 
with the fourth mass eigenstate as a virtual particle were also considered, the corrections to the scattering amplitude 
being negligible.
\end{abstract}


\maketitle
\section{Introduction}

The standard model yields a precise description of the fundamental interactions. Nevertheless, in the neutrino  sector there 
are still observed phenomena whose proper description requires the introduction of new elements. Indeed, on one side the tiny neutrino 
squared mass  differences suggests very small neutrino masses whose explanation requires the introduction of new heavy fields;
on the other side, we have not been able to explain the short-baseline anomalies such as the LSND signal \cite{lsnd}, 
the MiniBooNE excesses \cite{miniboone}, and the reactor anomaly \cite{reactor}. 

The standard picture includes 3 active neutrinos whose mixing is described by the Pontecorvo-Maki-Nakagawa-Sakata (PMNS) matrix. 
The mixing angles and mass differences have been measured in different experiments which are designed to maximize a desired 
effect depending on the source, baseline and energy. The last parameter to be measured was the  mixing angle $\theta_{13}$ 
\cite{dayabay1,reno}, that was for a long time assumed to be zero.

In this work we explore the consequences on CP asymmetries in neutrino oscillations due to the very existence of an additional 
heavy neutrino. As measured on LEP, the invisible width of the $Z$ boson imposes that an additional active neutrino must 
be heavier than $m_Z/2$, while LEP2 excluded a heavy charged lepton, that would be the doublet partner of the neutrino considered, 
up to $100$ GeV \cite{pdg}. The most attractive scheme would be to consider a fourth sequential chiral generation, and indeed, 
there are several motivations to consider this as a first attempt to modify the Standard Model (SM). These have already been 
summarized in \cite{holdom} and include: new CP Violation (CPV) source for the baryon asymmetry of the universe problem \cite{Hou:2008xd,Hou:2011df,Fok:2008yg,Ham:2004xh}, new perspectives in the Higgs naturalness problem or in the fermion mass 
hierarchy problem \cite{Holdom:1986rn,King:1990he,Hill:1990ge,Frampton:1999xi,Delepine:2010vw}. However,  the recent 
LHC (ATLAS and CMS) results put strong limits on the fourth generation Standard Model (SM4) since they
have excluded at 95\% a fourth generation down ($b'$) and up quark ($t'$) with masses smaller than $m_{b'}<611$ GeV \cite{Chatrchyan:2012yea,ATLAS:2012aw,Aaltonen:2011vr} and  $m_{t'}<557$ GeV \cite{CMS:2012ab,Aad:2012xc,Aad:2012bt,Abazov:2011vy}. 
Since these quarks couple to the Higgs boson with a strength proportional to its mass, they do not decouple from the production 
of the Higgs boson. The existence of these extra fermions, regardless of their mass, would imply a Higgs boson $M_H> 600$ GeV. Also precision electroweak observables constrain the difference of fourth generation quark and lepton masses \cite{kribs,Novikov:2002tk,Baak:2011ze}.

In this paper though, we will make a discussion of the neutrino sector considering a fourth active neutrino regardless of 
the model in which it is embedded. Our goal is to test  if the two extra CP violating phases that appear in the $4\times 4$ 
PMNS matrix can yield a sizable asymmetry in neutrino  oscillations, hence we scan the whole parameter space for these phases 
obtaining  those values that yields the maximum CP asymmetry in the different channels.
For clarity and simplicity, we will start with the results obtained for neutrinos propagating in vacuum. We point how the 
somewhat surprising result of having a measurable asymmetry in the survival channel $\nu_\mu \to \nu_\mu$ is obtained due to unitarity 
and kinematical constraints. Then we will proceed in an analogous analysis 
to the case of neutrinos propagating through matter for distances and energies that are comparable to current experiments. 
Finally, we will calculate the order of the corrections that could be induced in the scattering amplitude due to virtual effects 
of the heavy neutrino and its mixing with the light  ones.

Our paper is organized as follows: in the next section we discuss neutrino oscillations in vacuum, the kinematics of the heavy 
neutrino and its impact on neutrino CP asymmetries. In section III we calculate the same observables in matter. In Section IV 
we calculate the box diagrams contributing at the next order in perturbation theory. Our conclusions are given in Section V 
and we give some details of the calculations in an appendix. 

\section{Neutrinos in Vacuum}

The neutrino of flavour $\alpha$, $\nu_\alpha$, is by definition the one that is produced in the weak interactions with its 
charged lepton partner $W^+\longrightarrow l_\alpha^++\nu_\alpha$, where $\alpha= e, \mu$ o $ \tau$. These are the interaction 
eigenstates which are related to the propagating states $\nu_{i}$, $i=1,2,3$ through the PMNS matrix $U_{\alpha i}$ as
\be \label{neutrinostate}
\vert \nu_\alpha \rangle=\sum_i U_{\alpha i}\vert\nu_{i}\rangle.
\ee
The lagrangian for charged weak currents is given in terms of the propagating neutrino state as
\be
\mathcal{L}_{cc}=-\frac{g}{\sqrt{2}}\sum_{\alpha,i}\bar{l}_{L\alpha}\gamma_{\mu}U_{\alpha i}\nu_{Li}W^{\mu}_{+}+h.c.,
\ee
Several considerations are made in order to describe a neutrino state at a distance $L$ from the production point and at a 
selected neutrino energy $E$. First, it is assumed that the neutrino mass eigenstate propagates as a plane wave, then we take 
the $z-$axis along the neutrino direction and consider that the propagating states are ultra-relativistic, i.e., $E_i \gg m_i$.
Under these assumptions one can easily find the neutrino state after a time $t$ during which it travels a distance $z$ \cite{Kayser:1981ye,Giunti:1993se,Beuthe:2001rc}
\ba
\vert \nu_\alpha (t)\rangle&=&U_{\alpha1}  e^{-i\phi_1}\vert \nu_1\rangle +U_{\alpha2}  e^{-i\phi_2}\vert \nu_2\rangle \nn\\ 
&+& U_{\alpha3}  e^{-i\phi_3}\vert \nu_3\rangle,
\ea
where $\frac{m_i^2}{2E_i}z\equiv \phi_i $.
The probability for oscillation of initial flavour $\alpha$ to final flavour $\beta$ is given by  
$|\langle \nu_\beta\vert\nu_\alpha (t)\rangle|^{2}$ and a 
straightforward calculation yields 
\ba \label{nuetonumu} \nn
P_{\alpha\beta}&=&2 \mathcal{R} (U_{\alpha1}U^{*}_{\beta1}U^{*}_{\alpha2}U_{\beta 2}[e^{-i(\phi_1-\phi_2)}-1])\\
&+&2 \mathcal{R} (U_{\alpha1}U^{*}_{\beta1}U^{*}_{\alpha 3}U_{\beta 3}[e^{-i(\phi_1-\phi_3)}-1]) \nn \\ 
&+&2 \mathcal{R} (U_{\alpha2}U^{*}_{\beta2}U^{*}_{\alpha 3}U_{\beta 3}[e^{-i(\phi_2-\phi_3)}-1]),
\ea
where the relation $\vert  a+b+c\vert^2=\vert a \vert^2+\vert b \vert^2+\vert c \vert^2+ 2 \mathcal{R} (ab^*+ac^*+bc^*)$ and 
the unitarity of the PMNS matrix are used. Equation (\ref{nuetonumu}) can also be written in the compact form
\ba \label{versimt} \nn
P_{\alpha\beta}=&-&4 \sum_{i<j}\mathcal{R} (U_{\alpha i}U^{*}_{\beta i}U^{*}_{\alpha j}U_{\beta j}) \sin^2 \left(\frac{\phi_j-\phi_i}{2}\right)\\
&-&2 \sum_{i<j}\mathcal{I} (U_{\alpha i}U^{*}_{\beta i}U^{*}_{\alpha j}U_{\beta j}) \sin \left(\phi_j-\phi_i\right).
\ea
It is clear that the second term in Eq. (\ref{versimt}) in general will yield a difference in $P_{\alpha\beta}$ with respect 
to $P_{\bar{\alpha} \bar{\beta}}$ and this effect is driven by the phases in the PMNS matrix.  

\subsection{The PMNS$_{4\times 4}$ Matrix and the Input Parameters.}
A unitary  $n_g\times n_g$ matrix is parametrized by $n_g^2$ parameters, out of which $2 n_g-1$ phases may be reabsorbed by 
rephasing the fields. On the other hand, an orthogonal matrix of the same dimension can be parametrized by $n_g(n_g-1)/2$ angles. 
So a unitary complex $n\times n$ matrix will have $\tfrac{1}{2}(n_g-1)(n_g-2)$ physical phases. For $n_g=4$ we have 6 rotation 
angles and 3 phases. The PMNS$_{n_g\times n_g}$ matrix can be written using the parametrization proposed in 
\cite{botella,fritzsch} as
\ba \nn
U&=&R_{n_g-1,n_g}\tilde{R}_{n_g-2,n_g}\ldots \tilde{R}_{1,n_g} \cdot \ldots \cdot R_{k-1,k} \nn \\&\ldots& R_{23} \tilde{R}_{13}R_{12},
\ea
where $\tilde{R}_{ij}$ is a complex rotation matrix on the $ij$ axis and ${R}_{ij}$ is the rotation without the phase. In the case 
$n_{g}=4$ 
\ba \nonumber
U=&(&w_{34}(\theta_{34})\times w_{24}(\theta_{24},\varphi_3) \times w_{14}(\theta_{14},\varphi_2))\\ 
&\cdot& (w_{23}(\theta_{23})\times w_{13}(\theta_{13},\delta) \times w_{12}(\theta_{12})).
\ea
In the calculation of the CP asymmetries below we will use for the conventional angles and differences of squared masses the 
best fit points from \cite{t2k}. Although there are more recent reports on the value of $\theta_{13}$, i.e. \cite{reno,dayabay1,doublechooz,dayabay2}, the measured values are consistent with those of \cite{t2k} 
and we prefer to work with the set of parameters that were used in the fit in a single analysis that considers 
the three neutrinos. As can be seen from Fig. 6 in \cite{t2k}, in this long baseline neutrino experiment is not possible to 
disentangle the values of $\delta$ and $\theta_{13}$, so we decided to keep the best fit point which yields $\delta=0$. 

The upper bounds on the additional parameters due to the existence of a fourth neutrino are obtained from deviations of the 
unitarity of the PMNS matrix. In this concern, considering electroweak decays, such as, W decay, invisible Z decay, test of 
universality and rare decays, which lead to upper bounds for the product $(NN^\dag)_{\alpha\beta}=(HV(HV)^\dag)_{\alpha\beta}$, 
where $N$ is the non-unitary PMNS matrix composed of a hermitian ($H$) and a unitary ($V$) matrix, robust bounds for deviations 
of unitarity of the PMNS matrix were obtained in \cite{meloni} which updates previous studies in \cite{antusch,antusch2}. 
We use the results reported in \cite{meloni} to extract the upper bounds on the the additional angles due to the existence 
of a fourth neutrino. Relating these bounds to the 3 new angles, we obtained: 
$\theta_{14}<3\text{.}62^\circ$,  $\theta_{24}<2\text{.}29^\circ$ and  $\theta_{34}<4\text{.}21^\circ$. Notice that we have 
left the extra CP phases, $\varphi_2$ and $\varphi_3$, as free parameters and that from now on we will have a complex PMNS matrix. 
We use the input parameters shown in Table \ref{input}, with $\Delta m^2_{ij}\equiv m_i^2-m_j^2$.
For the Normal Hierarchy (NH) $\Delta m^2_{31}=\Delta m^2_{32}+\Delta m^2_{21}$, whereas for the inverted hierarchy (IH) we 
used instead $\Delta m^2_{31}=-2\text{.}4\times 10^{-3}$ eV$^2$ and $\theta_{13}=10.99^\circ$.
\begin{table}
\bc
\begin{tabular}{ccc} 
 Parameter & Value & Reference \\ \hline
$\theta_{12}$&$34\text{.}4^\circ$&\cite{t2k}\\
$\theta_{13}^*$&$9\text{.}68^\circ$&\cite{t2k}\\
$\theta_{23}$&$45^\circ$ &\cite{t2k}\\ 
$\delta^*$&$0^\circ$ &\cite{t2k}\\ \hline
$\Delta m^2_{21}$&$7\text{.}6\times 10^{-5}$ eV$^2$ &\cite{t2k}\\
$\Delta m^2_{32}$&$2\text{.}4\times 10^{-3}$ eV$^2$&\cite{t2k}\\ \hline
$\theta_{14}$&$<3\text{.}62^\circ$ &\\
 $\theta_{24}$&$<2\text{.}29^\circ$&\\
$\theta_{34}$&$<4\text{.}21^\circ$ &\\ \hline
$m_{4}$&$\approx$100 GeV &\\ \hline
\end{tabular}
\caption{Input parameters assuming the Normal Hierarchy (NH). Two variables marked with $^*$ are correlated and we take their 
best fit values.\label{input}}
\ec
\end{table}

\subsection{Neutrino Oscillations with an Extra Heavy Neutrino}

If a fourth neutrino exists, it is necessarily heavy with a mass well above the present energies of neutrino beams, thus 
the oscillation of a light neutrino ($\nu_{e},\nu_{\mu},\nu_{\tau}$) into a heavy neutrino denoted hereafter as $\nu_{E}$ 
is kinematically forbidden. In spite of this, the very existence of such neutrino implies the appearance of new phases in 
the PMNS matrix which manifest in observable CP violation effects yielding asymmetries in the probabilities for neutrino 
oscillations with respect to those of the antineutrinos. Furthermore, as we will show below, the interference with light 
neutrinos produces interesting effects and even in the diagonal channels (surviving probabilities) there can be an observable 
asymmetry due to the combined effects of the new phases, unitarity and the kinematics of the oscillations. 

We consider first the propagation of neutrinos. In general, the proper calculation of the neutrino state at a given time 
requires to solve the corresponding wave equation whose complete form is still unknown due to the lack of information on 
the nature of the mass term. However, as far as the neutrinos do not 
interact with other fields while traveling, Lorentz covariance allows to write the neutrino state as 
\be \label{td}
\psi(t)=\psi(0) e^{-ip\cdot x}=\psi(0) e^{-i(Et-\vert \mathbf{p}\vert L)}
\ee
and the specific form of the state $\psi (0)$ is not required beyond the fact that it coincides with the states produced in 
weak interactions. The weak eigenstates  are linear combinations of the neutrino mass eigenstates which satisfy the 
eigenvalue equation 
$H\vert \nu_a\rangle=E_a\vert \nu_a\rangle$, $a=1,2,3,4$, thus they have a component along the fourth neutrino 
\be
\vert\nu_\alpha \rangle=U_{\alpha1}\vert\nu_1\rangle+U_{\alpha2}\vert\nu_2\rangle+U_{\alpha3}\vert\nu_3\rangle
+ U_{\alpha4}\vert\nu_4\rangle
\ee 
The fourth component is rapidly damped by the kinematics during the propagation. Indeed, in contrast to the light neutrinos, 
a heavy one propagates non-relativistically and we can estimate the size of this component at a time $t$ by a simple quantum 
mechanical calculation. The kinematical suppression of this component during the propagations can be seen as an imaginary 
momentum according to
\be
\vert \mathbf{p}\vert \equiv -i\omega,  \qquad \omega=\sqrt{m^2-E^2}.
\ee
Performing a non-relativistic expansion it is straightforward to show that  
\begin{equation}
-i(Et-\vert \mathbf{p}\vert L)=-m L\left[1+\mathcal{O}\left(\frac{E}{m}\right)\right].
\end{equation}
We conclude that for the fourth heavy neutrino instead of a phase we have an exponential damping factor $e^{-m L}$ which kills this 
component after a distance of order $1/m$. For instance, considering a 100 GeV mass eigenstate this component disappears after a 
distance of the order of $ 10 ^{-3}$ fm. The propagation effects of a fourth heavy neutrino seem to be completely irrelevant. We 
remark however that due to the propagation of $|\nu_{i}\rangle $ and the mixing of the neutrinos, at a given time the state 
$|\nu (t)\rangle $ contains a non-vanishing component of the heavy weak eigenstate $|\nu_{E}\rangle $ and there is a non-vanishing 
probability for the oscillation to this state. Of course, at the present beam energies this oscillation is forbidden by the 
kinematics, but then the effects of this non-vanishing probability manifest via unitarity.   

In order to explore the possible effects we now calculate  the oscillation probabilities in the presence of a fourth neutrino. 
The generalization of Eq. (\ref{nuetonumu}) is straightforward
\ba   \label{probab4}
P_{\alpha\beta}&=&2\sum_{i<j} \mathcal{R} (U_{\alpha i}U^{*}_{\beta i}U^{*}_{\alpha j}U_{\beta  j}[e^{-i(\phi_i-\phi_j)}-1]) \\ \nn
&+& \sum_{i} 2 \mathcal{R}(U_{\alpha i}U^{*}_{\beta i}U^{*}_{\alpha 4}U_{\beta 4} (e^{-i\phi_i}e^{-mL}-1)).
\ea
Neglecting the term containing the damping factor we get the generalization of Eq. (\ref{versimt}) as  
\ba \label{versimt4} \nn
P_{\alpha\beta}=&-&4 \sum_{i<j}\mathcal{R} (U_{\alpha i}U^{*}_{\beta i}U^{*}_{\alpha j}U_{\beta j}) \sin^2 \left(\frac{\phi_j-\phi_i}{2}\right)\\
&-&2 \sum_{i<j}\mathcal{I} (U_{\alpha i}U^{*}_{\beta i}U^{*}_{\alpha j}U_{\beta j}) \sin \left(\phi_j-\phi_i\right) \\ \nn
&+& 2 \sum_{i}\mathcal{R} (U_{\alpha i}U^{*}_{\beta i}U^{*}_{\alpha 4}U_{\beta 4}).
\ea
There are three major modifications due to the existence of a fourth neutrino: i) the new phases modify the second term in 
Eq. (\ref{versimt4}) yielding new sources for CP violation and CP asymmetries; ii) the appearance of the last term in this equation; 
and iii) the modification of unitarity relations. In the following we explore the consequences of these modifications for the 
asymmetries in neutrino oscillations.

\subsection{Symmetry Transformations and the Oscillation Probabilities}
The symmetry transformations $\mathbf{T}$, $\mathbf{CP}$ y $\mathbf{CPT}$  map the following neutrino oscillations amplitudes
\ba 
\mathbf{T} &:& \nu_\alpha\rightarrow \nu_\beta \Longrightarrow  \nu_\beta\rightarrow \nu_\alpha\\ 
\mathbf{CP} &:& \nu_\alpha\rightarrow \nu_\beta \Longrightarrow  \bar{\nu}_\alpha\rightarrow \bar{\nu}_\beta\\ 
\mathbf{CPT} &:& \nu_\alpha\rightarrow \nu_\beta \Longrightarrow  \bar{\nu}_\beta\rightarrow \bar{\nu}_\alpha,
\ea
thus $\mathbf{CPT}$ symmetry requires
\be  \label{cpt} 
P_{\beta\alpha}=P_{\bar{\alpha} \bar{\beta}}.
\ee
On the other hand, exchanging $\nu_\alpha\leftrightarrow \nu_\beta$ in Eq. (\ref{versimt}), if the PMNS matrix is complex we get
\be \label{notem}
P_{\alpha \beta}\neq P_{\beta\alpha}.
\ee
This means that the weak interactions are not invariant under the $\mathbf{CP}$ (or $\mathbf{T}$) transformation. Combining the 
results in Eqs. (\ref{cpt},\ref{notem}), for a complex PMNS matrix we get
\be  
P_{\alpha\beta}\neq P_{\bar{\alpha}\bar{\beta}},
\ee
and in general it is possible to test $\mathbf{CP}$ symmetry in weak interactions by considering asymmetries in 
neutrino oscillations.

Concerning the surviving probabilities, $\mathbf{CPT}$ symmetry requires
\be
P_{\alpha\alpha}=P_{\bar{\alpha}\bar{\alpha}}
\ee
hence $\mathbf{CP}$ asymmetries should not show in the survival probabilities. However, the transitions to the heavy state are forbidden by kinematics and this manifests in observable oscillation asymmetries even in these channels. Indeed, probability conservation requires
\be \label{unitaryrelation}
P_{\alpha \alpha}=1- \sum_{\beta\neq\alpha}P_{\alpha \beta},
\ee
where $\beta=e,\mu,\tau,E$. Similarly
\be \label{unitarityanti}
P_{\bar\alpha \bar\alpha}=1- \sum_{\beta\neq\alpha}P_{\bar\alpha \bar\beta},
\ee
Thus, using both unitarity and $\mathbf{CPT}$ symmetry we get
\be
\sum_{\beta\neq\alpha}P_{\alpha \beta}=\sum_{\beta\neq\alpha}P_{\bar\alpha \bar\beta}
\ee
i.e., $\mathbf{CP}$ violation effects cancel in the sum over all the channels. 
However, in spite of being allowed by the dynamics, for present beam energies well below the heavy neutrino mass, the transition of a light 
neutrino into the heavy one is forbidden by the kinematics. On the other hand for a complex PMNS matrix from Eq. (\ref{versimt4}) 
we get
\be
P_{\alpha E}\neq P_{\bar\alpha \bar E}
\ee 
then the combined effect of kinematics, unitarity and CP violation is a measurable asymmetry in the survival channels even if 
$\mathbf{CPT}$ is a good symmetry, e.g. for the muon neutrinos, due to the kinematically forbidden transition with probabilities
 $P_{\alpha E}$ and $P_{\bar\alpha \bar E}$, in an experiment 
in general we will obtain 
\be
1-P_{\mu e}-P_{\mu \tau} \neq 1- P_{\bar{\mu}\bar{e}}- P_{\bar{\mu} \bar{\tau}}.
\ee

\subsection{CP Asymmetries}

In this section we study the size of the effects of a heavy neutrino in neutrino oscillations asymmetries as a function of the
involved parameters. The sizeable effects of 
a heavy neutrino in CP asymmetries were shown in  \cite{czakon} in the case where the extra phases are fixed to $\pi/2$. This part 
of our calculation updates the input and generalizes results in \cite{czakon} to scan the whole parameter space for the new phases and 
to keep the PMNS matrix elements corresponding to the fourth neutrino independent and within the maximal values allowed by unitarity. 

We start by fixing the beam energy and baseline to the MINOS (M) and T2K (T) type setup, leaving the new phases as free 
parameters and finding the values maximizing the effects. For $E$ we use the mean beam energy in both cases. The specific 
values used in the calculation are given in Table \ref{MT}. For each channel we obtain the values of the new phases maximizing 
the CP asymmetry
\be
A_{CP}=\frac{P_{-}}{P_{+}}, \qquad \text{ where } P_{\pm}=P_{\alpha \beta}\pm P_{\bar{\alpha} \bar{\beta}}.
\ee
The explicit expressions for the asymmetry in terms of the angles and phases characterizing the PMNS matrix are lengthy, so we 
only give in the appendix a simplified expression in the channel $\nu_\mu \to \nu_e$, where the effects of the new phases 
are visible. We will not consider at all times the last term in Eq. (\ref{versimt4}) since the modifications to the asymmetries due to it are negligible.

\begin{figure}[ht]
\begin{center}
\begin{tabular}{cc}
\begin{tabular}{|c|c|} \hline
\multicolumn{2}{|c|}{\large{$\nu_\mu\longrightarrow \nu_e$}} \\
NH&IH \\ \hline
\rule{0pt}{20.5ex}\scalebox{0.169}{\includegraphics{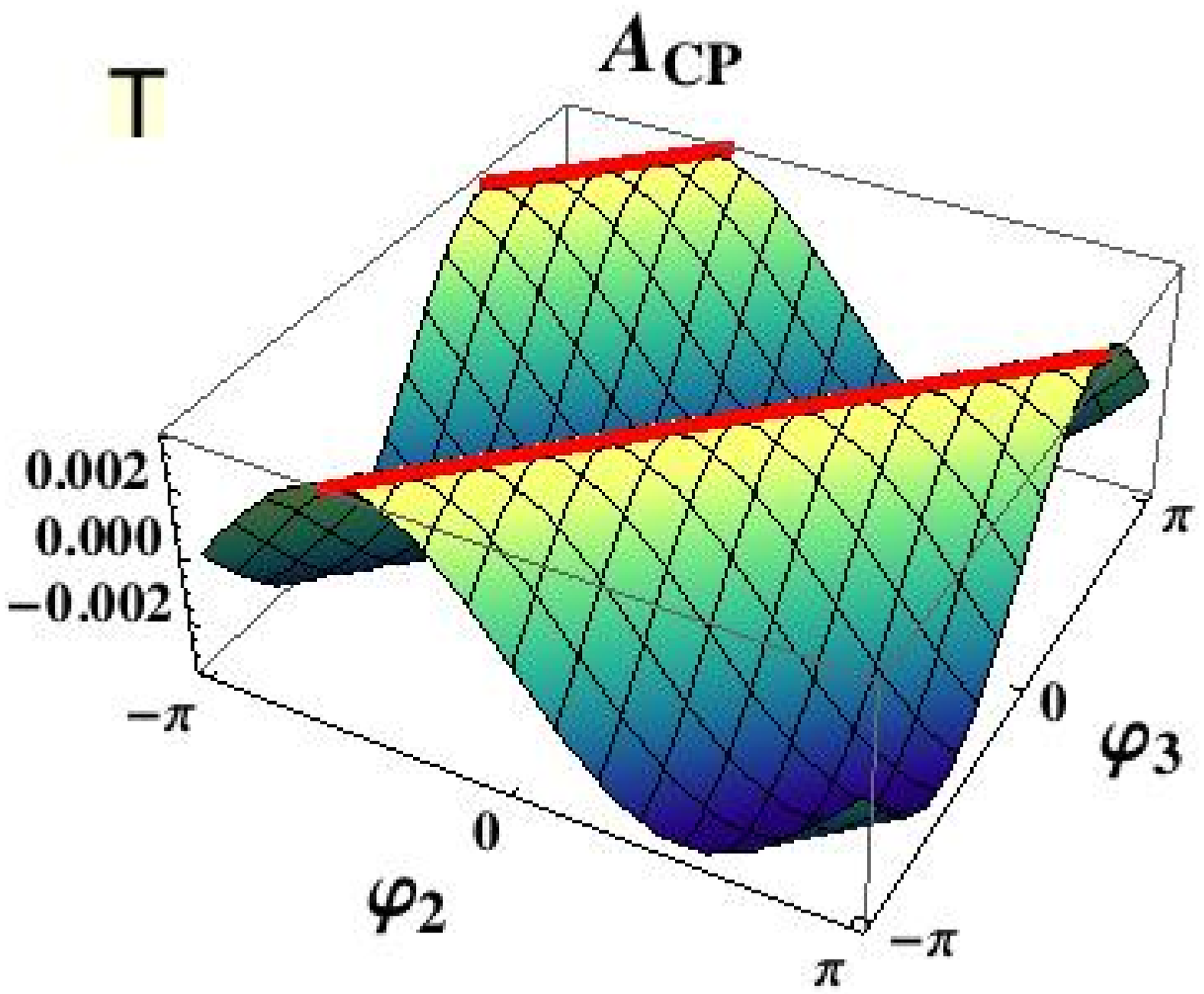}}& \scalebox{0.17}{\includegraphics{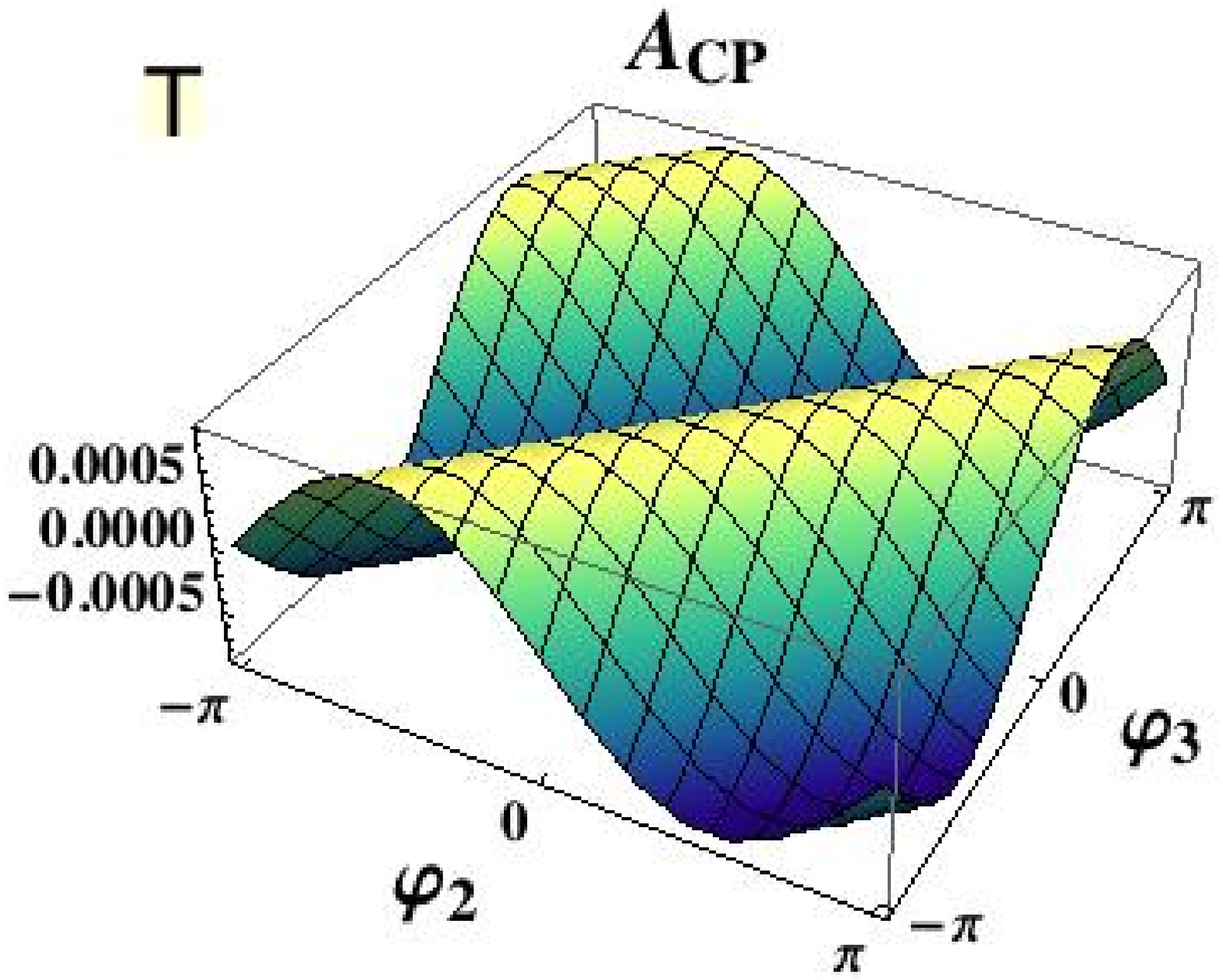}} \\ \hline
\rule{0pt}{20.5ex}\scalebox{0.17}{\includegraphics{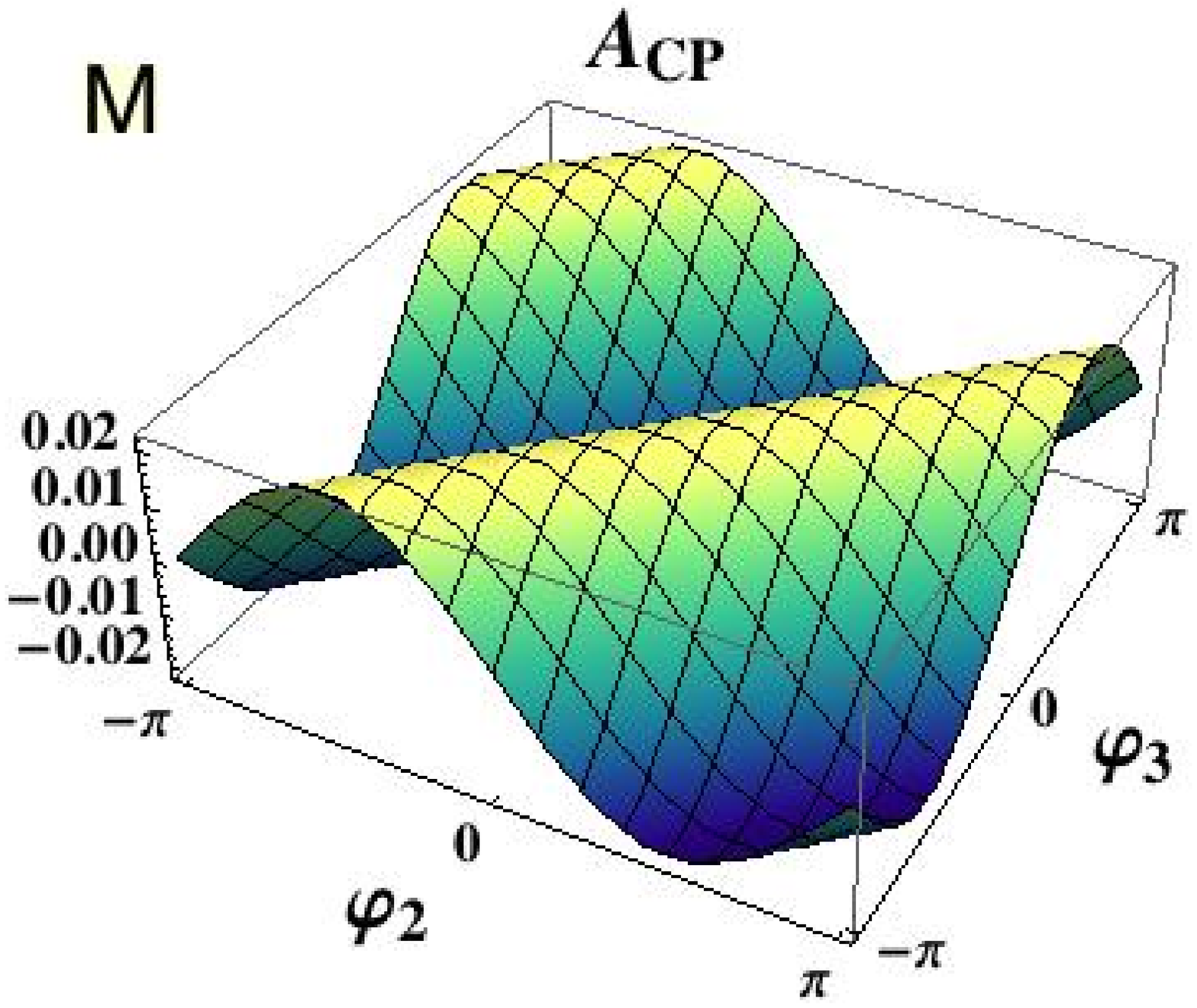}}& \scalebox{0.17}{\includegraphics{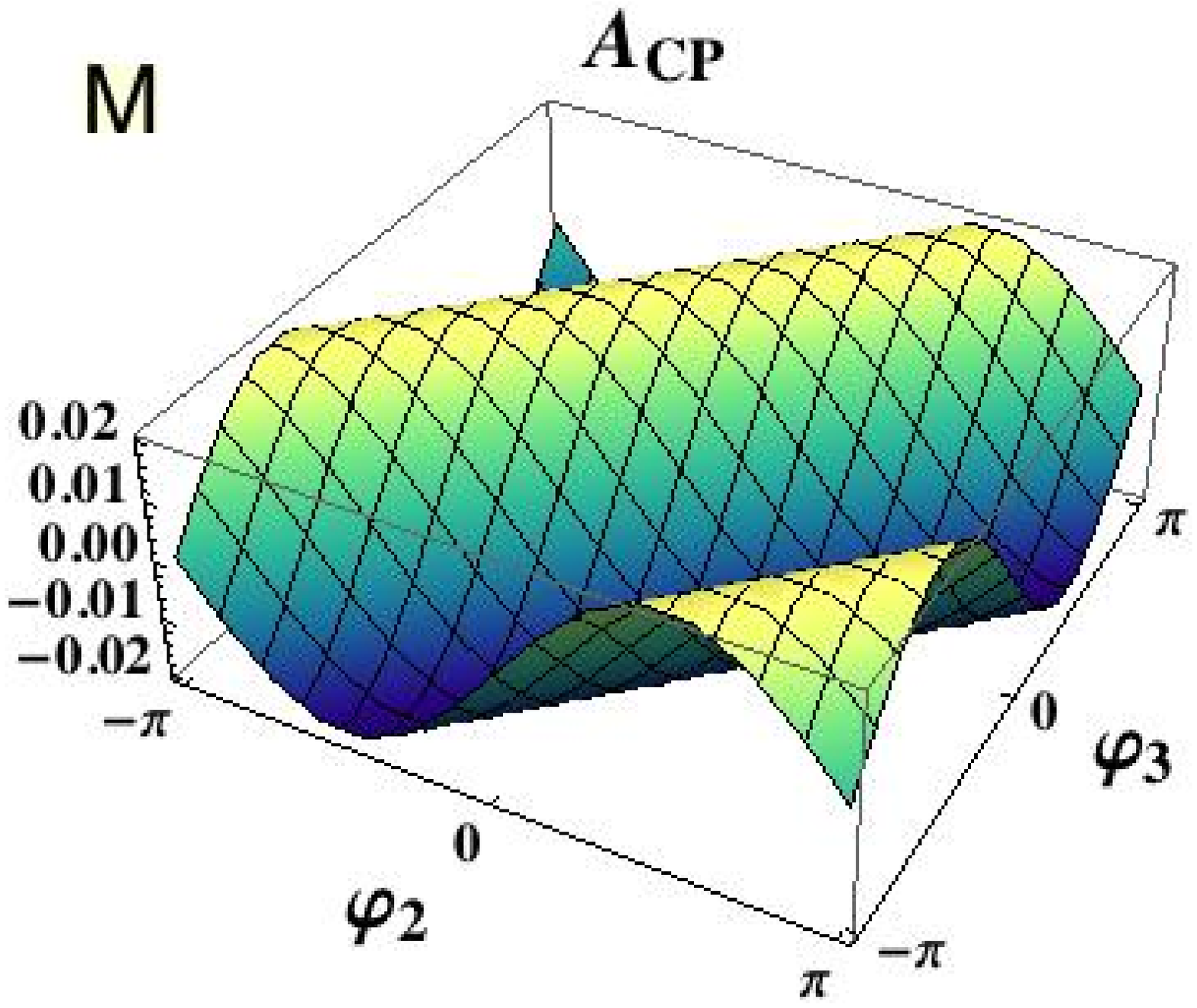}} \\ \hline \end{tabular} \\
\end{tabular}
\caption{ CP asymmetry in the $\nu_\mu\rightarrow \nu_e$ channel for the T and M parameters, considering the normal (NH) and inverted hierarchies (IH). \label{acpvacioT2K}}
\end{center}
\end{figure}

\begin{figure}[ht]
\begin{center}
\begin{tabular}{|c|c|}  \hline
\multicolumn{2}{|c|}{\large{$\nu_\mu\longrightarrow \nu_\mu$}} \\ 
NH&IH \\\hline
\rule{0pt}{20.5ex}\scalebox{0.17}{\includegraphics{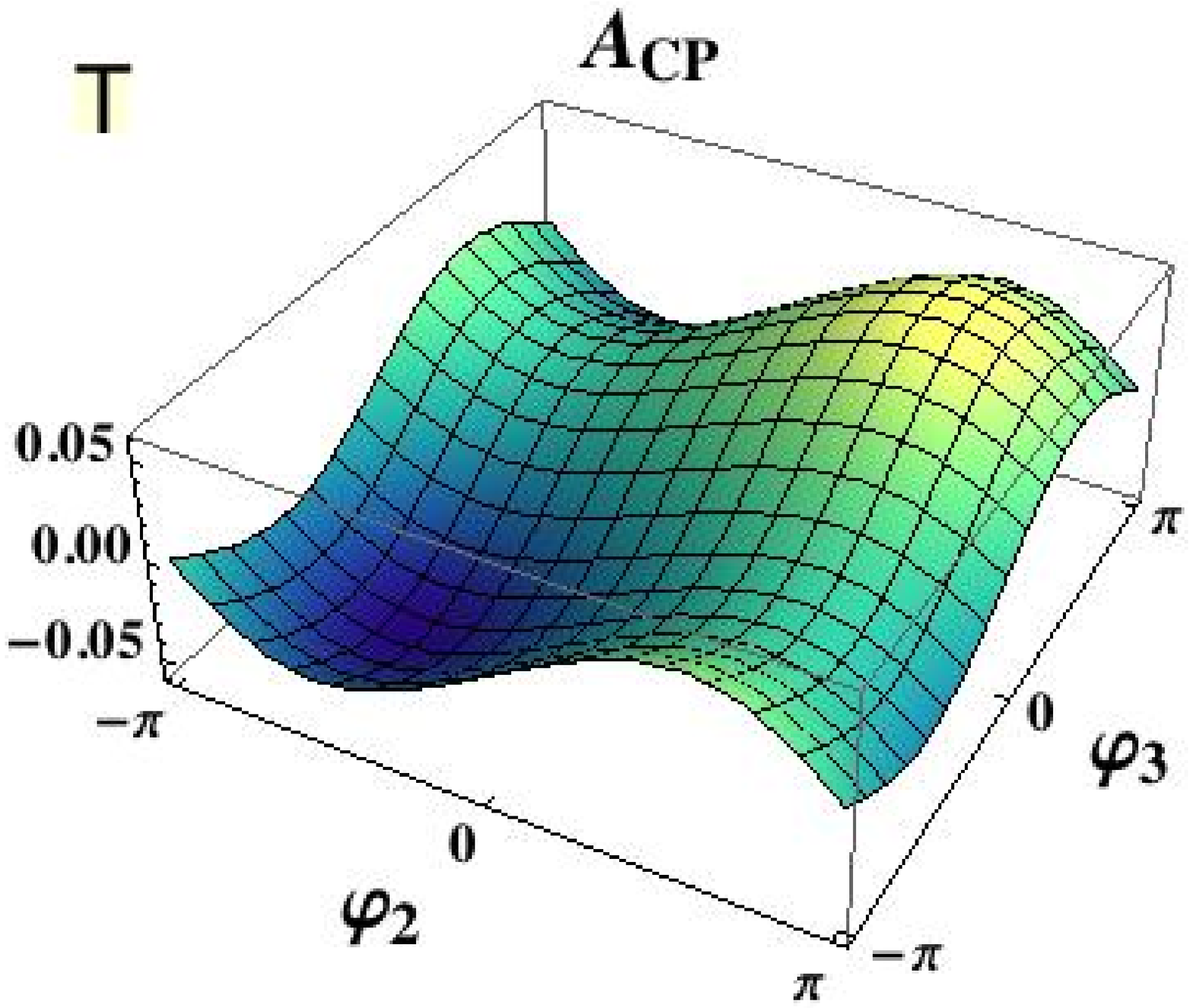}}& \scalebox{0.17}{\includegraphics{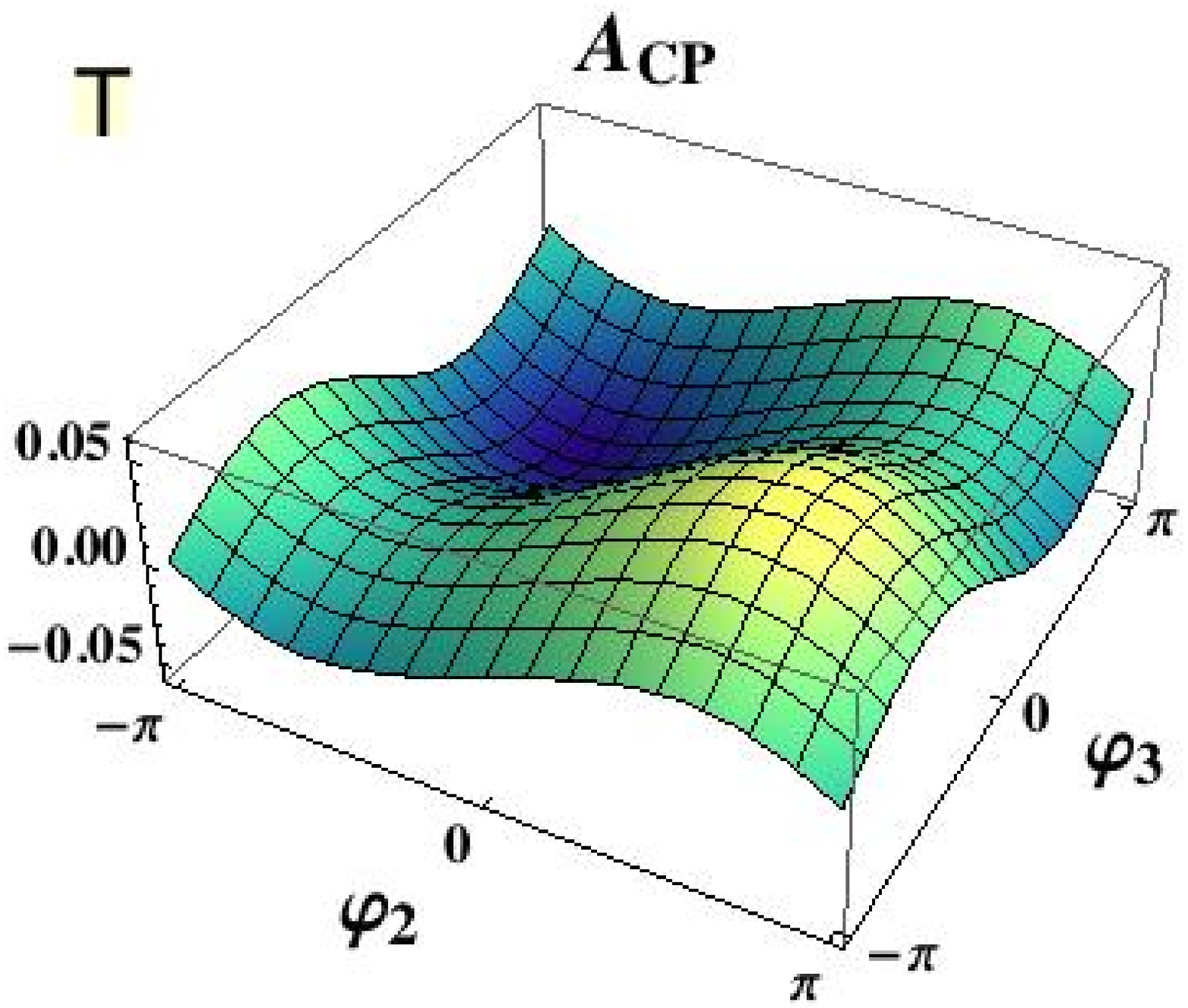}} \\ \hline
\rule{0pt}{20.5ex}\scalebox{0.17}{\includegraphics{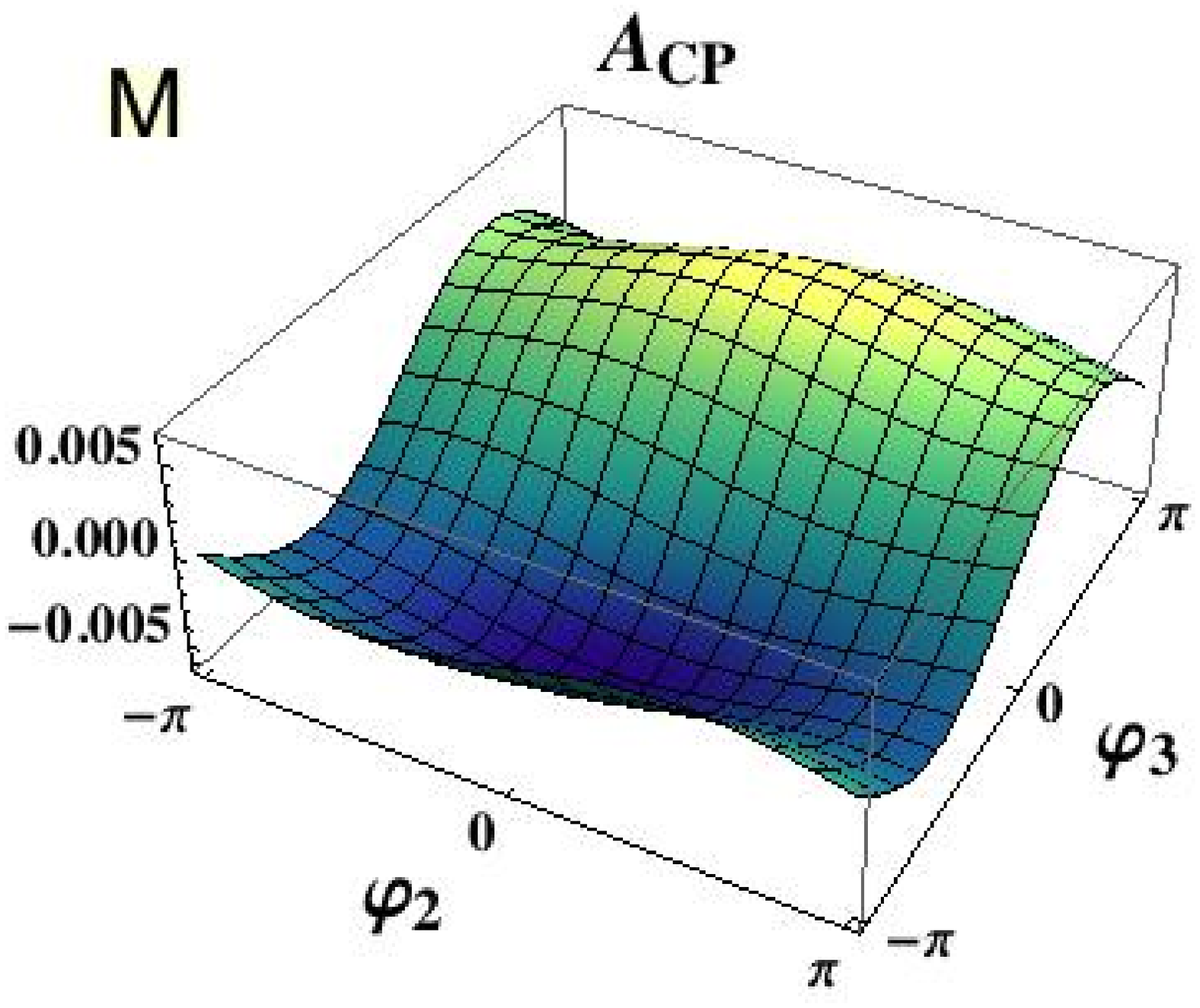}}& \scalebox{0.17}{\includegraphics{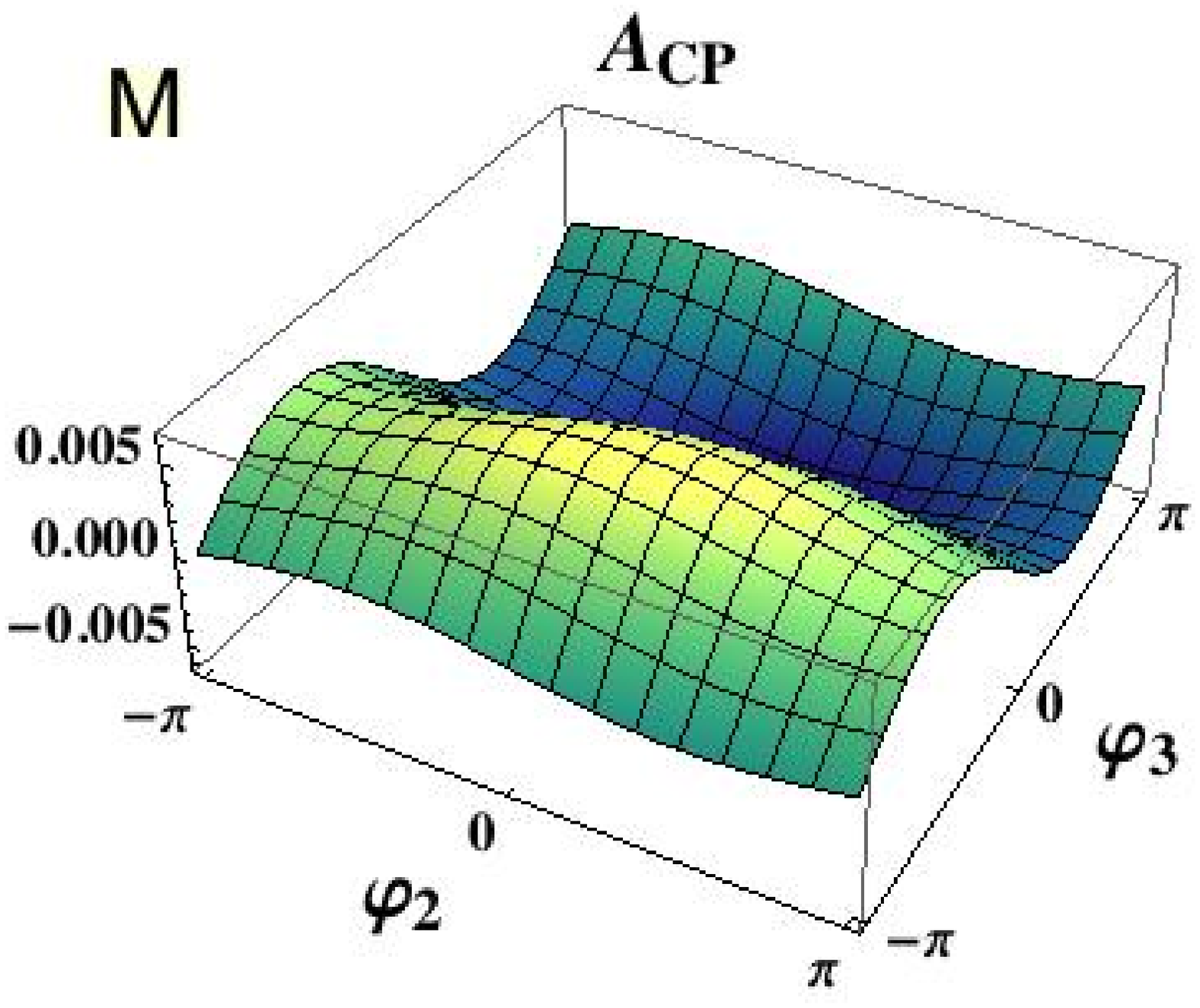}}   \\ \hline
\end{tabular}
\caption{CP asymmetry in the survival channel $\nu_\mu\rightarrow \nu_\mu$ for the T and M parameters, considering normal (NH) and 
inverted (IH) hierarchies. \label{acpvacioMINOS}}
\end{center}
\end{figure}

Our results for the CP asymmetry in the $\nu_\mu\to \nu_e$ oscillation channel are shown in Fig. \ref{acpvacioT2K} and those of the 
$\nu_\mu \to \nu_\mu$ survival channel are depicted in Fig. \ref{acpvacioMINOS}.
\bc
\begin{table}
\begin{tabular}{lccc}\\
Experiment & $L$ [km] & &$\langle E_\nu\rangle$ [GeV] \\ \hline
T2K (T) &295 &&0.6\\
MINOS (M) &735& &3.0 \\
\end{tabular}
\caption{Values used in the calculation of CP asymmetries}
\label{MT}
\end{table}
\ec

In these plots we can see the maximum values reached by the corresponding CP asymmetry and its dependence on the new phases. For 
the $\nu_\mu\to\nu_e$ channel there are regions in the parameter space yielding similar values for the maximum CP asymmetries as shown 
with a line in the upper left plot in Figure \ref{acpvacioT2K}. Also, it is remarkable that the survival $\nu_\mu\to \nu_\mu$ channel 
has the biggest asymmetry when evaluated for the T parameters. Comparing between the two hierarchies we find a 
maximum asymmetry 4 times bigger for the normal hierarchy compared with the inverted hierarchy in the $\nu_\mu\to \nu_e$ channel and for the 
T parameters. For the remaining channels we have a small variation for both, T and M parameters. We have studied all the channels 
finding the maximum asymmetries and the corresponding phases for each channel given in Table \ref{tablafases}. We find that the 
most favoured channels are the $\nu_\mu\to \nu_\mu$ survival channel for T-like setup and the $\nu_e\to \nu_\tau$ channel for the M-like setup which yield 
maximum asymmetries of the order of $5\%$.

\begin{table}[h!]
\begin{center}
\begin{tabular}{cc|cccc} \cline{1-3}
\multicolumn{3}{|c|}{\textcolor{azulin}{T-like}}&&&\\ \cline{1-3}
&&$\nu_\mu\rightarrow \nu_e$& $\nu_\mu\rightarrow \nu_\tau$&$\nu_e\rightarrow \nu_\tau$& $\nu_\mu\rightarrow \nu_\mu$\\  \hline 
&$\varphi_2$&$-93\text{.}66 ^\circ$&$-108\text{.}77 ^\circ$&$92\text{.}81 ^\circ$&$100\text{.}30 ^\circ$\\ 
NH&$\varphi_3$&$175\text{.}21 ^\circ$&$-70\text{.}07 ^\circ$&$-88\text{.}26 ^\circ$&$101\text{.}39 ^\circ$\\  
& $A_{CP}$&$0\text{.}3 \%$&$0\text{.}1 \%$&$0\text{.}5 \%$&$5\text{.}9 \%$\\  \hline 
&$\varphi_2$&$-35\text{.}09 ^\circ$&$-71\text{.}42 ^\circ$&$92\text{.}19 ^\circ$&$102\text{.}71 ^\circ$\\  
IH&$\varphi_3$&$-126\text{.}11 ^\circ$&$70\text{.}35 ^\circ$&$-88\text{.}06 ^\circ$&$75\text{.}80 ^\circ$\\  
&$A_{CP}$&$0\text{.}1 \%$&$0\text{.}1 \%$&$0\text{.}7 \%$&$5\text{.}7 \%$\\  \hline
\end{tabular}
\end{center}

\begin{center}
\begin{tabular}{cc|cccc} \cline{1-3}
\multicolumn{3}{|c|}{\textcolor{azulin}{M-like}}&&&\\ \cline{1-3}
&&$\nu_\mu\rightarrow \nu_e$& $\nu_\mu\rightarrow \nu_\tau$&$\nu_e\rightarrow \nu_\tau$& $\nu_\mu\rightarrow \nu_\mu$\\  \hline 
&$\varphi_2$&$-35\text{.}08 ^\circ$&$-110\text{.}42 ^\circ$&$-92\text{.}30 ^\circ$&$7\text{.}85 ^\circ$\\ 
NH&$\varphi_3$&$-126\text{.}09 ^\circ$&$-89\text{.}35 ^\circ$&$-81\text{.}77 ^\circ$&$91\text{.}52 ^\circ$\\ 
& $A_{CP}$&$2\text{.}2 \%$&$0\text{.}7 \%$&$4\text{.}3 \%$&$0\text{.}6 \%$\\  \hline 
&$\varphi_2$&$79\text{.}59 ^\circ$&$-65\text{.}38 ^\circ$&$91\text{.}80 ^\circ$&$8\text{.}68^\circ$\\ 
IH&$\varphi_3$&$170\text{.}75 ^\circ$&$89\text{.}49 ^\circ$&$-79\text{.}97 ^\circ$&$-88\text{.}44 ^\circ$\\  
&$A_{CP}$&$2\text{.}0 \%$&$0\text{.}7 \%$&$3\text{.}7 \%$&$0\text{.}6 \% $\\  \hline
\end{tabular}
\end{center}
\caption{Values of the CP asymmetry and phases $\varphi_2$ and $\varphi_3$ that maximize it in vacuum. \label{tablafases}}
 \end{table}

For given values of the new phases we get very different results for these setups and this lead us to study the asymmetries 
as a function of the baseline and of the neutrino beam energy. We are interested in the maximum asymmetry provided by the 
new phases, thus for a given value of $E$ and $L$ we scan the phases parameter space keeping those yielding the maximum asymmetry.

\begin{figure}[ht]
\begin{center}
\begin{tabular}{cc}
&\begin{tabular}{|c|} \hline$\nu_\mu\longrightarrow \nu_e$\\ \hline \end{tabular}\\
\scalebox{0.40}{\includegraphics{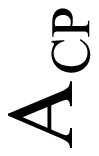}}&\scalebox{0.42}{\includegraphics{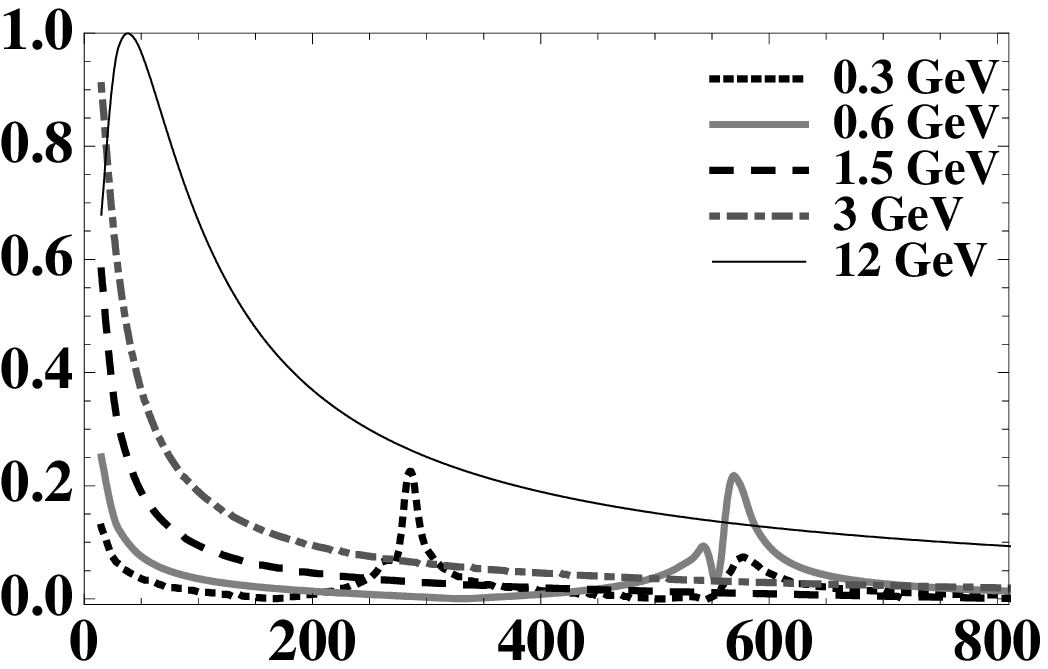}}\\
\end{tabular}
\begin{tabular}{cc}
&\begin{tabular}{|c|} \hline$\nu_\mu\longrightarrow \nu_\mu$\\ \hline \end{tabular}\\
\scalebox{0.40}{\includegraphics{fig9}}&\scalebox{0.42}{\includegraphics{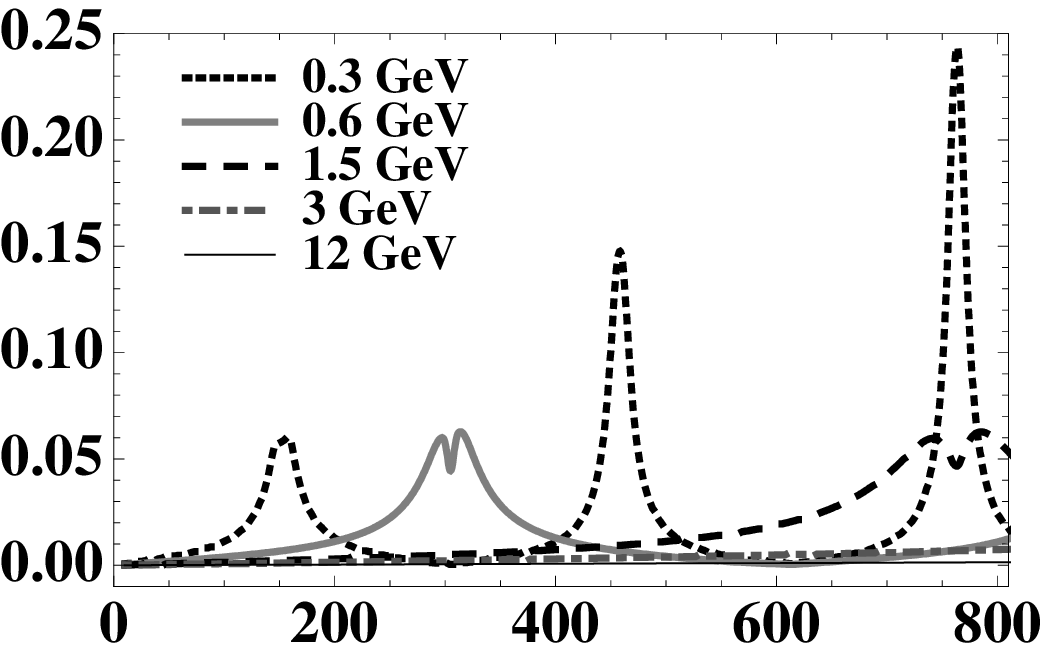}}\\
 &Baseline [Km]\\
\end{tabular}
\caption{Maximum CP asymmetry in the $\nu_\mu\rightarrow \nu_e$ and $\nu_\mu\rightarrow \nu_\mu$ channels as a function of the 
baseline considering NH. The curves correspond to selected beam energies from 0.3 GeV to 12 GeV. 
\label{VacioLNH}}
\end{center}
\end{figure}

\begin{figure}[ht]
\begin{center}
\begin{tabular}{cc}
&\begin{tabular}{|c|} \hline$\nu_\mu\longrightarrow \nu_e$\\ \hline \end{tabular}\\
\scalebox{0.40}{\includegraphics{fig9}}&\scalebox{0.42}{\includegraphics{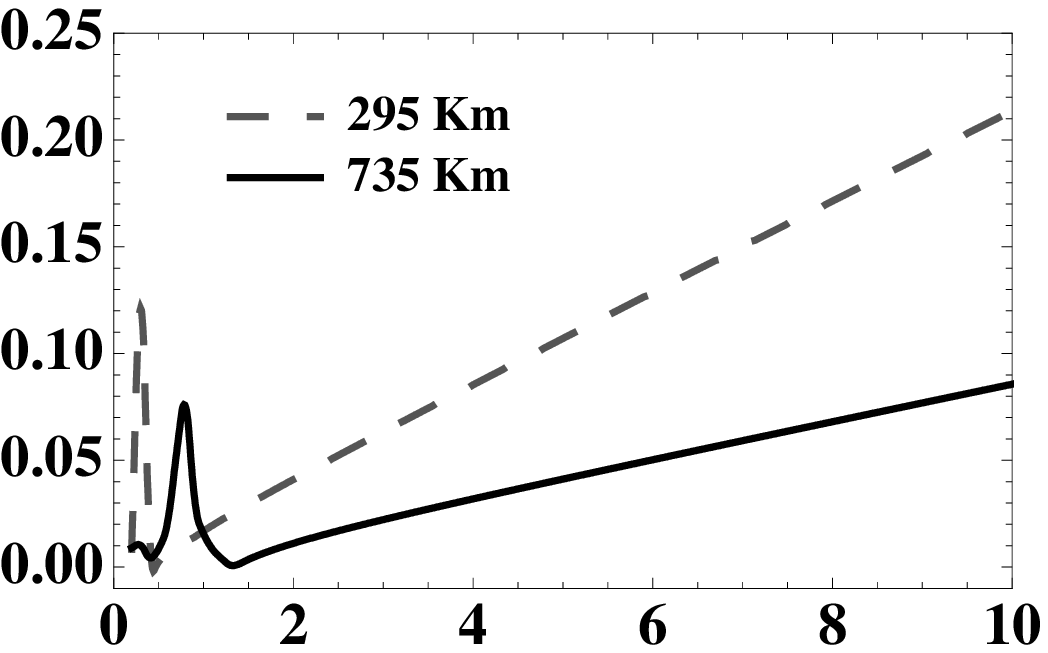}} \\
\end{tabular}
\begin{tabular}{cc}
&\begin{tabular}{|c|} \hline$\nu_\mu\longrightarrow \nu_\mu$\\ \hline \end{tabular}\\
\scalebox{0.40}{\includegraphics{fig9}}& \scalebox{0.42}{\includegraphics{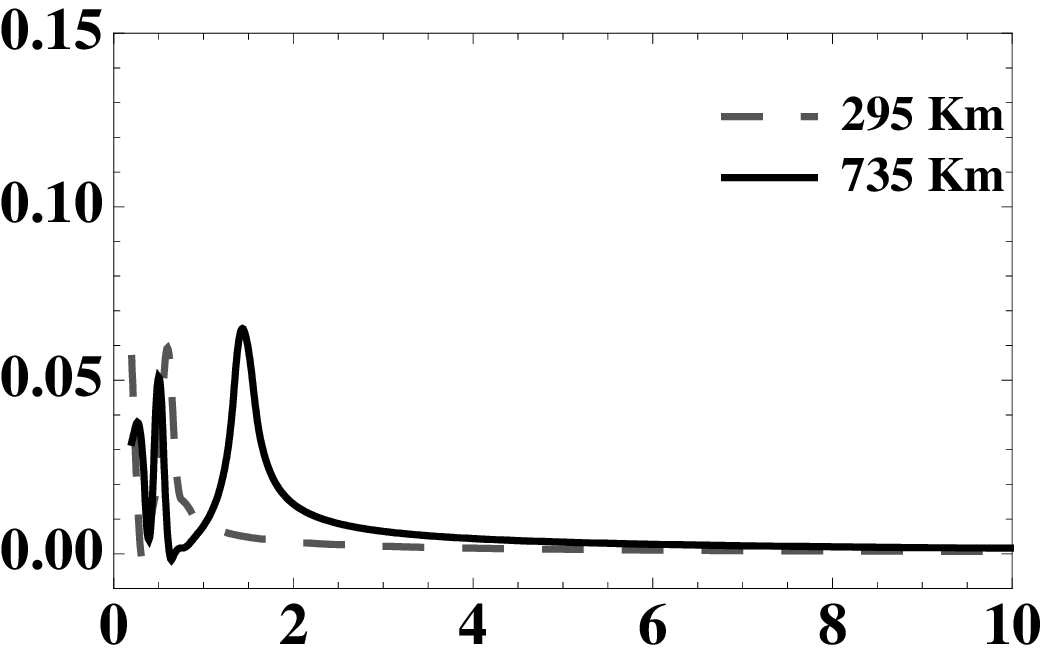}} \\
& Energy [GeV]\\
\end{tabular}
\caption{Maximum CP asymmetry as a function of the neutrino's energy for a baseline of 295 Km and  735 Km considering the NH.}
\label{VacioENH}
\end{center}
\end{figure}

Our results for the maximum asymmetry as a function of $L$ and $E$ and floating phases $\varphi_{2},~\varphi_{3}$ are given in Figs. \ref{VacioLNH} and \ref{VacioENH} respectively where we consider normal hierarchy.  

\begin{figure}
\bc
\begin{tabular}{cc}
\scalebox{0.40}{\includegraphics{fig9}}& \scalebox{0.4}{\includegraphics{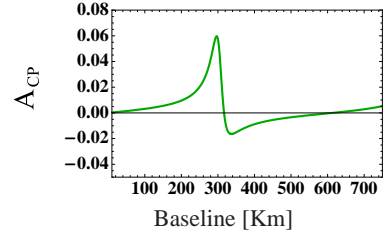}}\\
& Baseline [Km] \\
\end{tabular}
\caption{CP Asymmetry for fixed $\varphi_2$ and $\varphi_3$ using the T energy of 0.6 GeV as a function of the baseline 
with the NH.}
\label{base}
\ec
\end{figure}

In order to visualize the way the asymmetry behaves as a function of the baseline once the phases are fixed, we consider 
the T-like beam energy and evaluate in the phases corresponding to the maximum for the T setup. Our results for the $\nu_\mu \to \nu_\mu$ 
channel with normal hierarchy are given in Fig. \ref{base} were a variation of the order of $6\%$ can be seen when varying 
the baseline. 

From these plots we conclude that high energy neutrinos may give a big asymmetry for shorter baseline experiments in the 
appearance channels $\nu_\mu \to \nu_e$, $\nu_\mu\to\nu_\tau$ and $\nu_e \to \nu_\tau$. CP asymmetries can be substantial for low energy neutrinos only for 
specific values of the baselines. We studied these observables using the inverted hierarchy obtaining similar results. It 
is interesting that sizeable asymmetries in the survival channel 
$\nu_\mu \to \nu_\mu$, can be obtained for low energy neutrinos, for example with a baseline between 700 and 800 km and $E_{\nu}$ around 
0.3 GeV, while there is almost no CP asymmetry when considering higher energy neutrinos.  

\section{Neutrinos Through Matter}

As originally pointed out in \cite{wolfens,Barger:1980tf,Mikheev:1986if}, there might be non negligible effects on neutrino 
oscillations due to the interactions with matter in which the neutrinos propagate. The interaction between neutrinos and matter 
components: protons, neutrons and electrons, is feeble but a considerable effect on the oscillation amplitude can be obtained 
by coherent forward scattering of neutrinos from many particles in the material medium.

\begin{figure}[ht]
\begin{center}
\scalebox{0.74}{\includegraphics{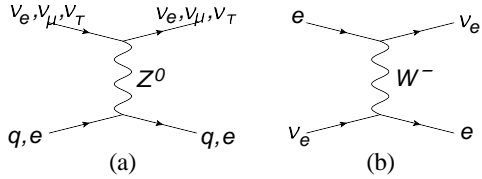}}
\caption{Tree level diagrams for neutrino interactions with particles in matter.  \label{tree}}
\end{center}
\end{figure}

At tree level, the neutral current (Z exchange in Fig. \ref{tree}(a)) yields an identical contribution for all neutrino 
flavours and it only produces a shift in the energy eigenvalue $E_i$. This shift does not affect the oscillation probabilities, 
since it appears as an overall phase factor in the amplitude. The charged current breaks this picture since matter contains 
only electrons and at tree level the only process contributing is $\nu_e e$ elastic scattering via the 
$u$-channel $W$ exchange diagram shown in Figure \ref{tree}b. At the present beam 
energies well below the $W$ mass this diagram yields the following effective interaction  
\ba \label{heff}
H_{eff}&=&\frac{G_F}{\sqrt{2}}\bar{e}\gamma^\mu(1-\gamma_5)\nu \bar{\nu}\gamma_{\mu}(1-\gamma_5)e   \label{heff}\\ \nonumber
&=&\frac{G_F}{\sqrt{2}}\bar{e}\gamma^\mu(1-\gamma_5)e \bar{\nu}\gamma_{\mu}(1-\gamma_5)\nu
\ea
where a Fierz transformation has been performed to obtain the result on the second line. With $N_e \approx 1\text{.}5 N_A/\text{cm}^3$, considering a constant density $\rho \approx 3 \text{ g}/\text{cm}^3$, the final correction to the energy is $\sqrt{2}G_F N_e$.

Neutrino propagation in matter requires to take into account these interactions and the time evolution of an arbitrary state vector $\vert \nu (t)\rangle =\sum_i f_i(t) \vert \nu_i \rangle$ that will be driven by the equation \cite{wolfens,Barger:1980tf,Mikheev:1986if}
\ba \label{bargereq5}
i \frac{d  f_j (t) }{dt}&=&\frac{m_j^2}{2E} f_j (t)  + \sum_\alpha \sqrt{2}G_F N_e U_{e j}^*U_{e k} f_k (t)  \nn\\
&\equiv&H_{jk} f_k (t).
\ea
In \cite{Barger:1980tf} an iterative  solution was found. The leading order for $f_j^{(i)}$ $(i=1,...,n)$  are 
\be
f_j^{(i)}(t=0)=\delta_{ij}.
\ee
Then, assembling the row vectors into a $n\times n$ matrix $X$ that satisfies the equation
\be \label{hx}
i\frac{dX}{dt}=XH,
\ee
with the boundary condition $X(t=0)=1$ an analytical solution for (\ref{hx}) is possible for a constant electron density $N_e$. 
 This approximation is valid if we restrict to $L<750$ km.

\begin{figure}
\begin{center}
\begin{tabular}{|c|c|} \hline
\multicolumn{2}{|c|}{NH}\\ 
$\nu_\mu\longrightarrow \nu_e$&$\nu_e\longrightarrow \nu_\tau$ \\ \hline
\rule{0pt}{20.5ex}\scalebox{0.17}{\includegraphics{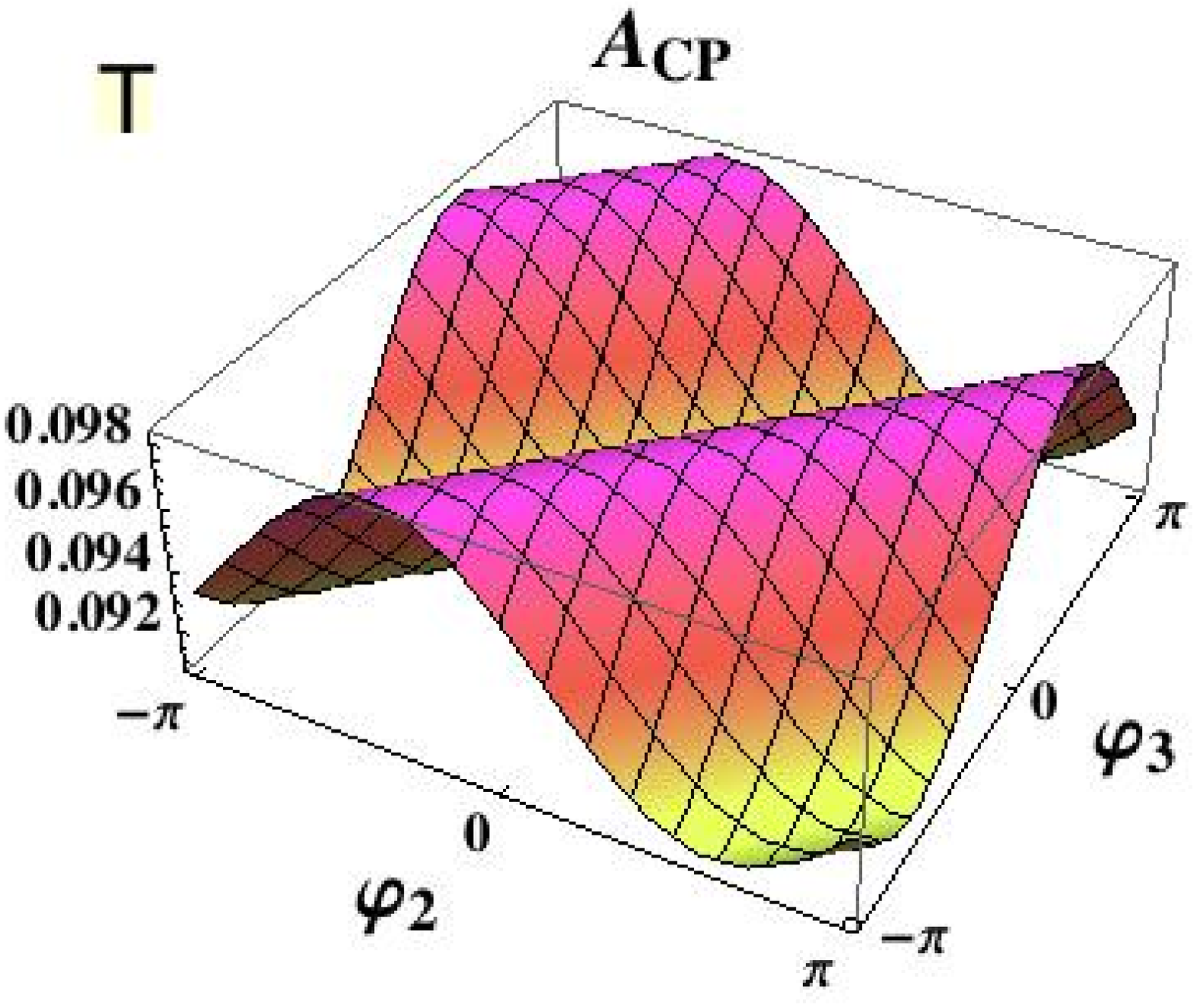}}& \scalebox{0.17}{\includegraphics{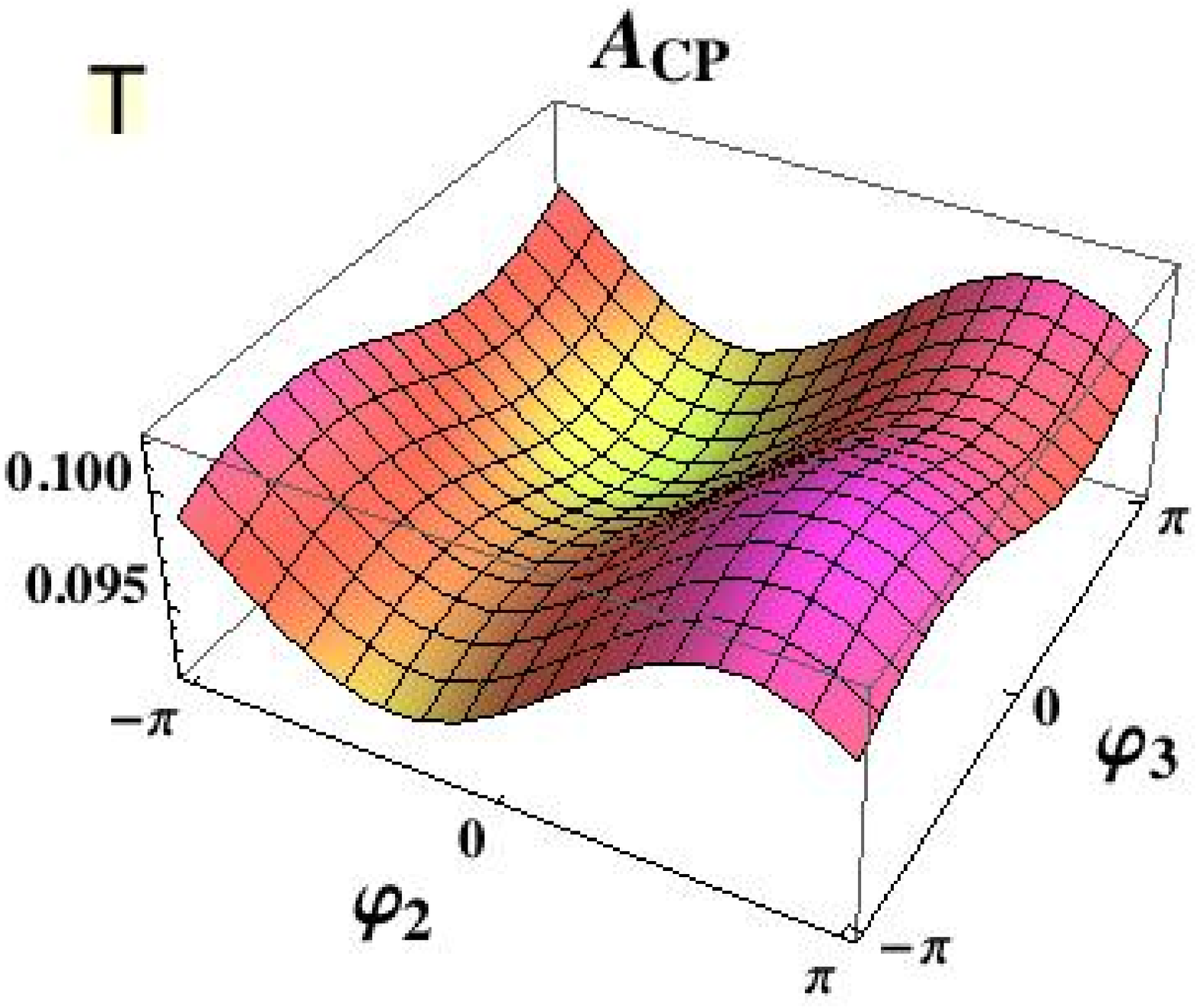}} \\  \hline
\multicolumn{2}{c}{(a)}\\
\multicolumn{2}{c}{}\\
\end{tabular}

\begin{tabular}{|c|c|} \hline
\multicolumn{2}{|c|}{\large{$\nu_\mu\longrightarrow \nu_\mu$}}\\ 
NH&IH \\ \hline
\rule{0pt}{20.5ex}\scalebox{0.17}{\includegraphics{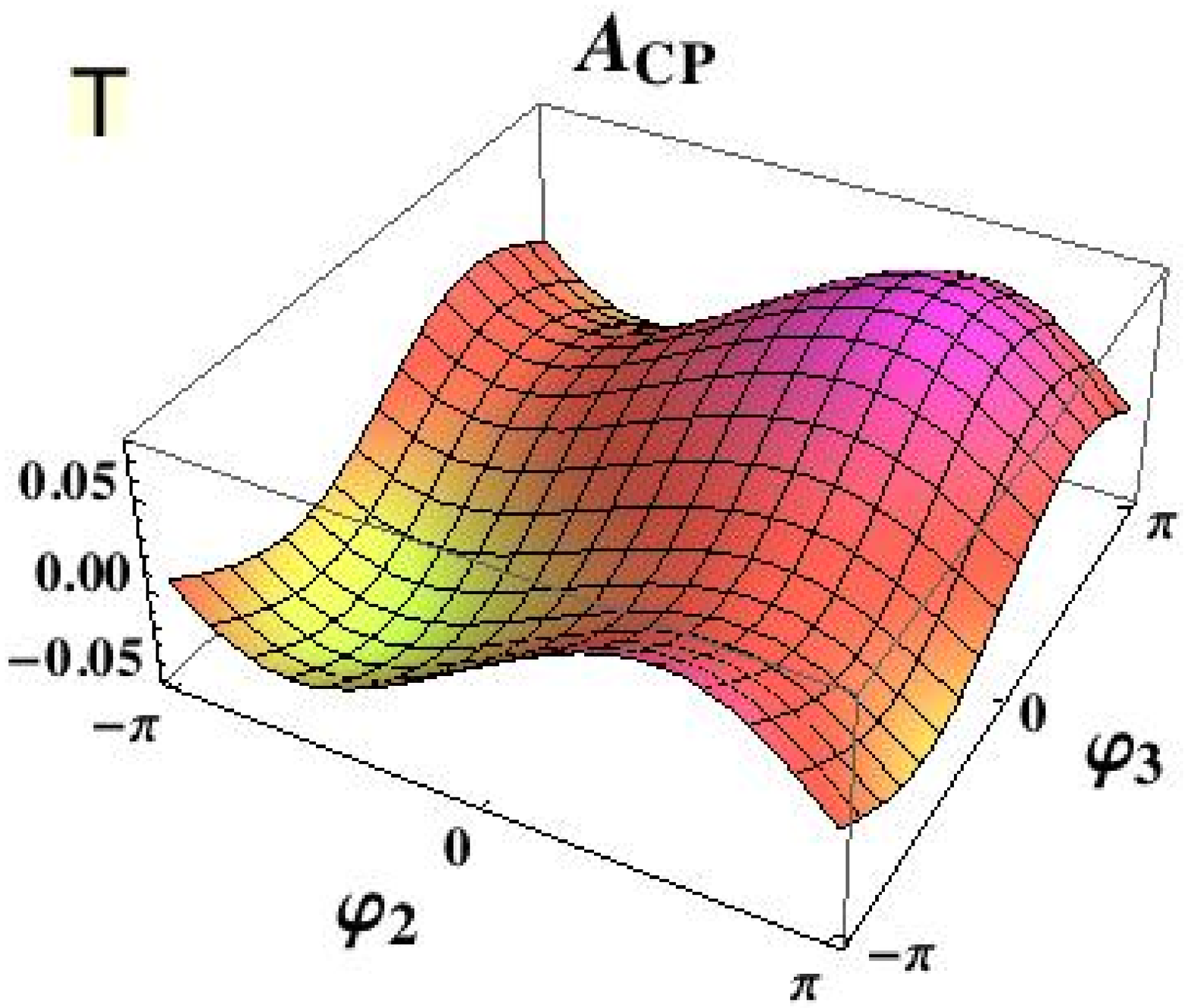}}& \scalebox{0.17}{\includegraphics{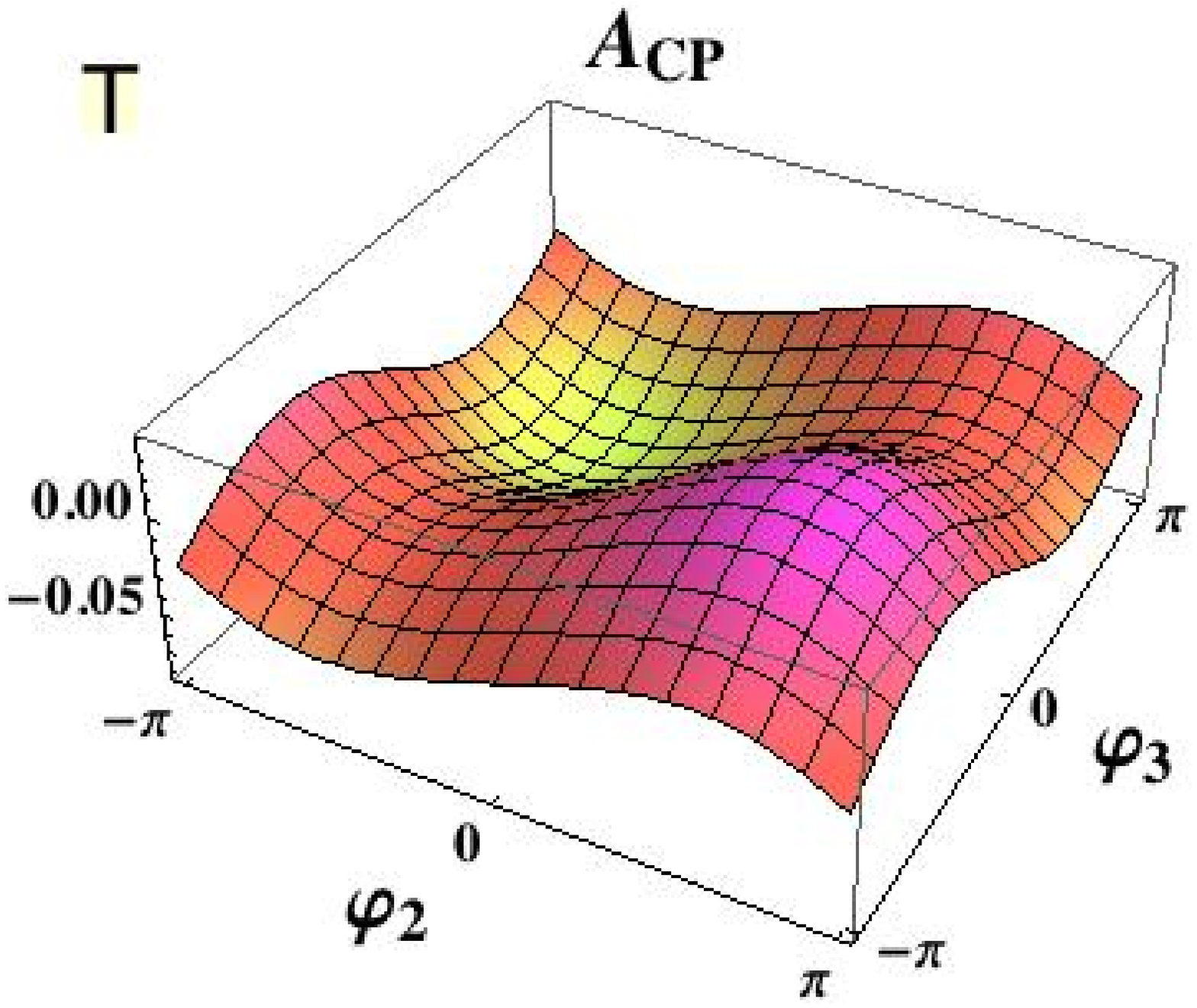}} \\  \hline
\multicolumn{2}{c}{(b)}\\
\end{tabular}

\caption{CP asymmetry in the (a ) $\nu_\mu \rightarrow \nu_e$ and the $\nu_e \rightarrow \nu_\tau$ channels with NH (b) $\nu_\mu \rightarrow \nu_\mu$ channel for the normal (NH) 
and inverted (IH) hierarchy for the T-like experiment parameters considering matter effects. \label{acpmateriaT2K}}
\end{center}
\end{figure}

\begin{figure}
\begin{center}
\begin{tabular}{cc}
&\begin{tabular}{|c|} \hline$\nu_\mu\longrightarrow \nu_e$ \\ \hline \end{tabular}\\
&\scalebox{0.44}{\includegraphics{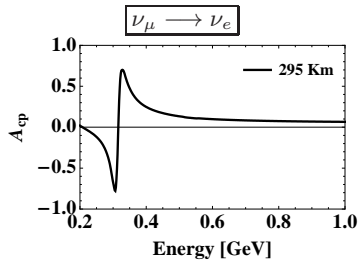}} \\
\end{tabular}

\caption{CP asymmetry in the $\nu_\mu\rightarrow \nu_e$ channel with low energy neutrinos considering matter effects. \label{materiamue}}
\end{center}
\end{figure}

We use the results in \cite{Barger:1980tf} and scan the parameter space for the new phases to obtain the maximum CP 
asymmetry in a T-like experiment. Our results are shown in Figs. \ref{acpmateriaT2K}, \ref{materiamue}. The asymmetries 
in general are bigger for the channels that involve $\nu_e$ as expected. While the results 
for the other channels show the same dependence on the neutrino energy compared to the vacuum case, in this channel 
the low energy neutrinos show an enhancement  providing a considerable asymmetry. In the survival channel a variation that depends on the hierarchy is found for the low energy neutrinos with respect to the vacuum result. 

\section{Box Diagrams with Virtual Heavy Neutrino}

\begin{figure}[!h]
\begin{center}
\scalebox{0.58}{\includegraphics{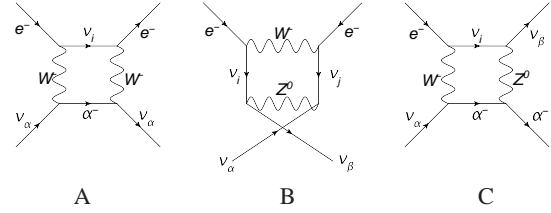}} \label{cajita}\\
\begin{tabular}{ccc} 
A \hspace{0.90in}& \hspace{0.90in}B& \hspace{0.90in}C  \\ 
\end{tabular}
\caption{Box diagrams with a virtual heavy neutrino for neutrino propagation through matter. \label{BOXfig}}
\end{center}
\end{figure}

Although box diagrams are suppressed with respect to the tree level contributions, enhancements of the matter effects can occur when 
we consider a heavy neutrino \cite{botella2}. Fig. \ref{BOXfig}a shows a box diagram that 
has already been calculated in \cite{botella2}, along with other one-loop diagrams representing the whole 
$\mathcal{O}(\alpha m_\tau^2/M_W^2)$ electroweak radiative corrections to coherent forward neutrino scattering. 
Here we do not consider the whole set of diagrams, since we are interested only in diagrams with a virtual heavy neutrino, therefore 
we focus on the box diagrams. We also consider as a possibility the diagrams shown on 
Figures \ref{BOXfig}b and \ref{BOXfig}c, where we can have neutrino flavour changing weak interactions since the virtual 
propagating neutrino is a mass eigenstate, thus we have the possibility of having these processes with different neutrino flavours 
as outer legs. 

The evaluations were performed using dimensional regularization and  the Feynman gauge ($\xi=1$). We considered the 
zero external momentum approximation, which is a good approximation for neutrino energies below 20 GeV. 

The diagrams with the corresponding Goldstone bosons turn out to be negligible, since they depend on powers of neutrino 
masses of the first three generations. For simplicity we stick to the normal hierarchy with the mass eigenstate $\nu_4$ being 
the most massive and used the lower bound on $m_{\nu_{4}}=100$ GeV. We would expect this one to be mostly $\nu_E$ 
with little components of the three standard flavours, since the mixing angles are very small. 
We used for $\varphi_2$ and $\varphi_3$, that appear in the relevant PMNS matrix elements, the values that maximize the 
CP asymmetry of the T parameters for the NH in each of the channels. It is only in the case of the flavour changing process 
that the extra CP violating phases enter into the expressions. In summary our results are the following:

A) For this  diagram the most relevant contribution to the amplitude comes from the light neutrinos $\nu_{i},~i=1,2,3$ as 
intermediate states. For an incoming electronic neutrino the absolute value of the amplitude from this diagram is of order 
$2 \times 10^{-7}$. The absolute value for the tree level amplitudes is $0.4  \times 10^{-4}$,  hence the ratio to the tree 
level diagram is of order $5\times 10^{-3}$. Amplitudes with incoming $\nu_{\mu}$, $\nu_{\tau}$ are even smaller.

B) This diagram allows for  neutrino flavour changing processes. For an incoming electronic neutrino and flavour conserving diagrams
 the ratio to the tree level diagram is of the order of $8 \times 10^{-4}$ while for flavour violating diagrams the absolute value of the amplitude is of the 
order of $ 10^{-10}$. Considering an incoming muonic neutrino with a final 
neutrino either muonic or taonic the order is even smaller, namely  $\sim 10^{-12}$. 

C) This diagram admits also change in neutrino flavour. For an incoming electronic neutrino the ratio to the tree level process 
in the $ee$  is $\sim 2\times 10^{-3}$, amplitudes for other flavour conserving processes and  amplitudes for flavour 
violating processes are at least two orders of magnitude smaller.

Summarizing this section we conclude that the amplitudes of the three box diagrams are at least 3 orders of magnitude smaller 
than the tree level amplitude and numerically have no impact in the asymmetries in neutrino oscillations. 

\section{Conclusions}

The very existence of a fourth neutrino modifies  the conventional PMNS mixing matrix introducing three new angles and two new 
phases yielding new possible sources for CP violation. Since the mixing with light neutrinos is small it is desirable to have an 
estimate of the size of the CP violation effects in current neutrino experiment.  

In this work we studied the maximum values that CP asymmetries in neutrino oscillations can reach when a fourth neutrino is considered, 
both in vacuum and in the presence of matter. We show that another consequence of the existence of a fourth neutrino is that it is 
possible to observe asymmetries in the surviving channels even if CPT is conserved, due to a combined effect of unitarity, kinematics 
and CP violation.

In the numerics we set the value of the phase appearing in the case of three neutrinos to $\delta=0$, so that the imaginary part 
of the PMNS matrix is due entirely to the existence of the fourth neutrino. 
Using the upper bound values for the three extra mixing angles as obtained form the deviations from unitarity studies \cite{meloni}, 
we got $\theta_{14}<3\text{.}62^\circ$,  $\theta_{24}<2\text{.}29^\circ$ and  $\theta_{34}<4\text{.}21^\circ$. We study the maximum 
asymmetries that can be reach in neutrino oscillation as functions of the neutrino energy, the distance to the detector and the two 
new phases, scanning the whole parameter space for the latter.

Our analysis shows that, under the above assumptions, in vacuum, the maximum asymmetries that can be reached in a T2K-like setup 
($L=295$ km,~$E_{\nu}=0.6$ GeV) are of the order of of $6\%$ and appear in the survival  $\nu_{\mu}\to\nu_{\mu}$ channel. 
Asymmetries in other channels are at least one order of magnitude smaller.  As for the MINOS-like setup 
($L=735$ km,~$E_{\nu}=3$ GeV), the maximum asymmetries can be reached in the $\nu_{e}\rightarrow\nu_{\tau}$ 
and $\nu_{\mu}\to\nu_{e}$ channel and are of the order of $4\%$ and $2\%$ respectively.
Importantly, for the same baseline but a different energy within the reach of MINOS, namely $E_{\nu}=1.4$ GeV, we get a maximum 
asymmetry of the order of $6\%$ in the surviving $\nu_{\mu}\to\nu_{\mu}$ channel.   

Effects of matter enhance the maximal asymmetry in the channels involving electronic neutrinos as expected. For the T2K-like setup
the maximum asymmetry in the $\nu_{\mu}\to\nu_{e}$ channel raises from $0.3\%$ in vacuum to  $9 \%$ in the presence of matter,
while for $\nu_{e}\rightarrow\nu_{\tau}$  it grows from $0.5\%$ in vacuum to  $10 \%$  in a material medium. In these channels results 
for the maximal asymmetries are similar with normal and inverted hierarchy. The interaction with matter has a small an indirect effect in the 
survival $\nu_{\mu}\to\nu_{\mu}$ channel via unitarity, whose sign depends on the hierarchy. It goes from the order of
$6\%$ in vacuum, independently of the hierarchy,  to $8\%$ in matter for normal hierarchy, while it decreases to $4\%$ in the case of 
an inverted hierarchy.  

The value of the asymmetry in the $\nu_{\mu}\to\nu_{e}$ channel varies with the energy of the neutrino beam in general being larger 
at low energies. For the T2K baseline, as shown in Fig. (\ref{materiamue}) the maximum asymmetries can be of the order of $70\%$ 
changing the sign when going from beam energies of $E_{\nu}=0.30$ GeV to $E_{\nu}=0.33$ GeV.  Although not shown  
explicitly, for a fixed energy below 1 GeV we obtain similar peaks in the maximum asymmetries for definite values of $L$. For the T2K 
energy $E_{\nu}=0.6$ GeV the peaks appear around $ L=550$ km.

We do not report the obtained asymmetries in matter in the case of MINOS-like because in this case the baseline is at the edge of the validity 
of the approximations used in the solution of the propagating equations and a deeper analysis beyond the scope of this work is necessary 
in this case. 

The potential contributions to the $\nu e$ scattering amplitude due to the exchange of virtual heavy neutrinos in box diagrams were found 
to be negligible.  

Finally we would like to remark that the possibility to have an asymmetry in the surviving channel $\nu_{\mu}\to\nu_{\mu}$ is a consequence 
of the very existence of a fourth neutrino, thus it would be interesting to try an experimental search of this channel since the measurement 
of an asymmetry in this channel would be a direct proof of the existence of a fourth neutrino with important implications in particle physics 
and cosmology.

\acknowledgments
Work supported by CONACyT M\'exico under project 156618.

\appendix
\section{CP Asymmetry for $\nu_\mu\rightarrow \nu_e$}
In order to sketch the way to obtain the asymmetry we will consider the $\nu_\mu \rightarrow \nu_e$ channel with some simplifying assumptions, namely that $\sin^2 2\theta_{13}=0$, in contrast with the assumptions in the rest of this work. As a notation we will simplify $s_{ij}=\sin\theta_{ij}$, $c_{ij}=\cos\theta_{ij}$ and $\Delta_{ij}=\Delta m^2_{ij} L/(4 E)$. In this case the probability for particle and antiparticle oscillations are given by
\begin{widetext}
\ba
P_{\mu e (\bar{\mu} \bar{e})}&=& s^2_{24}s^2_{14}c^2_{14}(c_{12}^4+s^4_{12}) +2 c^2_{24}c^2_{14}c^2_{23}s^2_{12}c^2_{12}(1-\cos \Delta_{21})\\ \nn
&+&2s_{24}c_{24}s_{14}c^2_{14}c_{23}s_{12}c^3_{12} \cos(\varphi_2-\varphi_3)-2s_{24}c_{24}s_{14}c^2_{14}c_{23}s^3_{12}c_{12} \cos(\varphi_2-\varphi_3) \\ \nn
&+&2s_{24}^2s^2_{14}c^2_{14}s^2_{12}c^2_{12}\cos \Delta_{21}\pm2s_{24}c_{24}s_{14}c^2_{14}c_{23}s_{12}c_{12}\sin(\varphi_2-\varphi_3)\sin  2 \Delta_{21}  \\ \nn
&-&2s_{24}c_{24}s_{14}c^2_{14}c_{23}s_{12}c^3_{12} \cos(\varphi_2-\varphi_3)\cos \Delta_{21}+2s_{24}c_{24}s_{14}c^2_{14}c_{23}s^3_{12}c_{12} \cos(\varphi_2-\varphi_3)\cos \Delta_{21}
\ea
 where we neglect the last term in Eq. (\ref{versimt4}). The difference in the probabilities is given in this case as
 \be
 P_{-}=4s_{24}c_{24}s_{14}c^2_{14}c_{23}s_{12}c_{12}\sin(\varphi_2-\varphi_3)\sin \Delta_{21} 
 \ee
\end{widetext}

\end{document}